# The Cyclotactor : Towards a Tactile Platform for Musical Interaction

Staas de Jong
LIACS
Leiden University

staas@liacs.nl

## ABSTRACT
This paper reports on work in progress on a finger-based tactile I/O device for musical interaction. Central to the device is the ability to set up cyclical relationships between tactile input and output. A direct practical application of this to musical interaction is given, using the idea to multiplex two degrees of freedom on a single tactile loop.

## 1. INTRODUCTION
In general, systems with haptic I/O will allow input to directly influence output. For example, a system might link pressing a button directly to some tactile vibration being output by a tactor (a small tactile stimulator, described in [4]). However, the opposite case, of output directly influencing input, is not necessarily possible. Where it is, cyclical relationships between tactile input and output can be set up: a certain tactile input may trigger some output response, which in turn influences the input, which again changes the output, and so on. Since this idea underlies the current device, it has been named *cyclotactor*.

The device is a work in progress, based on the prototype presented in [2]. The hardware of this prototype consisted of three main components: an electromagnet, a proximity sensor, and a handheld permanent magnet. The latter served both as the object of proximity sensing and as the transducer for electromagnetic force feedback, which was generated over a vertical distance range above the freely approachable device surface. The prototype was conceived specifically as a musical controller, and in order to illustrate this an example of its programmable behaviour was given, controlling a percussive sound.

The prototype was then further developed, resulting in a number of changes to its hardware components. A temperature sensor has been attached to the electromagnet component, while the device surface has been made vertically adjustable so as to precisely coincide with the distance at which magnetic output will be strongest. The electronics for proximity sensing have been replaced by a new circuit based on the reflection of infrared light, removing the previously existing dependence on environment light.

The handheld magnet has been replaced as well, by a combined permanent magnet / infrared reflector which is attached to a single finger by a velcro strap. This allows for easier interaction, since there is no need to hold a small object anymore, while the strap ensures that the magnet does not fly off during interaction. Also, all fingers are now potentially freed up to have their own I/O loops. The new component was tentatively named *keystone*, a word taken from architecture where it is used to denote the central supporting stone at the top of an arch, closing it.

The vertical finger movement supported by the resulting setup (see Figure 1) was intended to resemble that of playing a piano key [3], which in turn is a type of movement that few will dispute has proven itself in musical expression. Touching on a distinction stressed in [1], we also note that the force feedback can both have a kinaesthetic component (with force output influencing finger position) and a cutaneous component (with force output generating a vibration sensed by the skin).

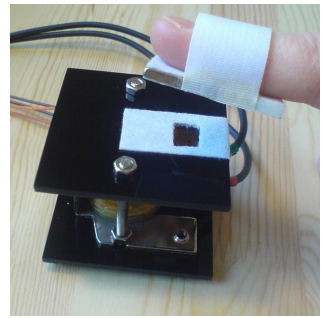

**Figure 1.** *The tactile interface.*

Effort was spent on improving the quality of tactile I/O. Gaining accurate magnetic output was only possible after implementing a robust and transparent mechanism for temperature compensation. This was verified to produce linear and identical output across the 20 - 100 °C operational temperature range. Output resolution was increased from 26 to at least 2195 steps. Although the output sample rate did not yet increase above 200 Hz, jitter reduction did improve timing precision.

Proximity input was made accurate and linear as well. In a reconfigurable trade-off, its sensitivity was fixed at 0.2 mm, with an associated distance range of 17 mm. The input sampling rate was increased from 100 Hz to 400 Hz. Finally, a range of reflective and magnetic materials were tested for use in the keystone component.

The next Section will aim to illustrate how the distinction made above between tactile display and cyclotactor can be an interesting one to make. It will give a direct practical application to musical interaction, using the idea to multiplex two degrees of freedom on a single tactile loop. The paper then ends with conclusions and future work.

## 2. MULTIPLEXING TWO DEGREES OF FREEDOM ON A SINGLE TACTILE LOOP

In the Introduction, the ability to set up cyclical relationships between tactile input and output was emphasized as a central defining feature of the prototype device. One such relationship could be based on the output of a fixed tactile wave. On the one hand, increasing finger proximity would increase the amplitude of the corresponding tactile vibration. (This is inherent to the type of magnetic field used.) On the other hand, the tactile vibration would in turn change finger proximity.

A closer look at this loop reveals another factor influencing the amplitude of the induced vibration: finger rigidity. Depending on its stiffness, a finger will variably dampen the vibration it is subject to (see Figure 2). Based on this, a prototype interaction was implemented which multiplexes two degrees of freedom on the proximity input. It does so by first averaging proximity over time in order to eliminate the influence of the output vibration cycle. This is illustrated in Figure 3, where the modified parameter has been named *nearness*.

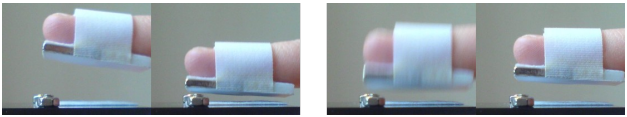

**Figure 2.** *Varying finger nearness (left) and rigidity (right).*

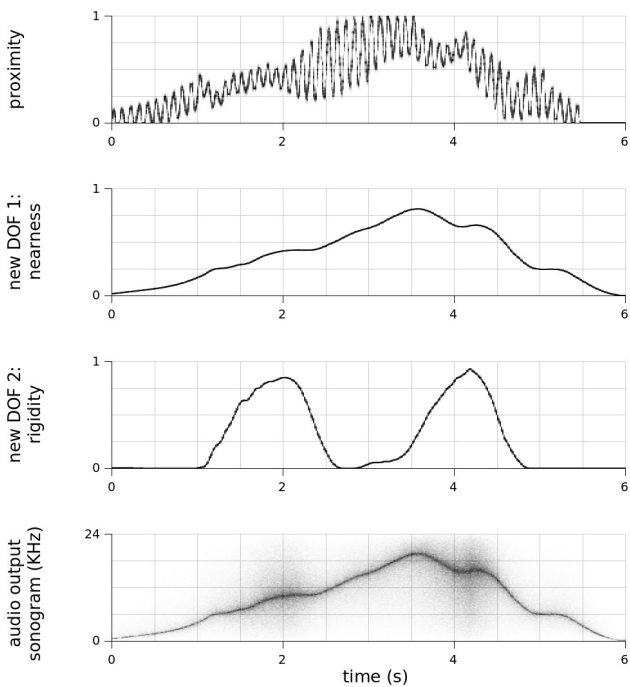

**Figure 3.** *I/O recorded during 6 seconds of interaction.*

Then, as a measure of the current tactile vibration amplitude, the deviation of proximity from its nearness average is also tracked and averaged over time. Before this can be used to determine finger rigidity, the effect the distance from the electromagnet has on vibration amplitude must be compensated for. This was done by measuring the typical vibration amplitude at a number of points across the nearness range, both while holding a finger maximally rigid and while holding it maximally loose. The resulting data was then used to implement a normalized *rigidity* parameter, orthogonal to nearness.

In order to provide a simple demonstration of exploring a two-dimensional sound space using the newly implemented degrees of freedom, both were connected to a single noise source, with nearness controlling its center frequency and rigidity its bandwidth. In Figure 3, below the two peaks in finger rigidity recorded during the slower movement in finger nearness, both parameters can be seen reflected in the sonogram of the system's audio output.

## 3. CONCLUSIONS AND FUTURE WORK

There are still technical issues hindering the cyclotactor's use as a truly open-ended platform for musical interaction. In [6], it was stated that for satisfying musical interfaces, the latency from sensor input to audio output should be 10 ms or less – and the current prototype certainly adheres to this. However, the latency from sensor input to tactile output should be 1 to 2 ms, if the kinaesthetic sense of touching a surface is to be adequately recreated [5]. This means that both the tactile output rate and the analog response time of the electromagnet must be improved. Another reason making this imperative is that the Nyquist frequency of the current setup is still well below the 250 Hz range at which the cutaneous sense is reported to be the most sensitive to vibrotactile feedback [4, 1].

Still, the improvements to control and feedback summarized in Section 1 have already enhanced considerably the device's ability to support intricately linked audio and tactile synthesis. The combined gains in accuracy, linearity and timing allowed implementation of Section 2's prototype interaction, intended to demonstrate the usefulness of emphasizing the distinction between tactile display and cyclotactor. Using the device's ability to set up cyclical relationships between tactile input and output, two virtual degrees of freedom were implemented on top of a single physical one.